# Mesoscopic 2D Charge Transport in Commonplace PEDOT:PSS Films

Y. Honma,[1] K. Itoh,[1,*] H. Masunaga,[2] A. Fujiwara,[3] T. Nishizaki,[4] S. Iguchi,[1] and T. Sasaki[1]

[1]*Institute for Materials Research, Tohoku University, Sendai 980-8577, Japan*
[2]*Japan Synchrotron Radiation Research Institute (JASRI)/SPring-8, Hyogo 679-5198, Japan*
[3]*Department of Nanotechnology for Sustainable Energy, Kwansei Gakuin University, Hyogo 669-1337, Japan*
[4]*Department of Electrical Engineering, Kyushu Sangyo University, Fukuoka 813-8503, Japan*

E-mail: k.itoh@imr.tohoku.ac.jp

**The correlation between the transport properties and structural degrees of freedom of conducting polymers is a central concern in both practical applications and scientific research. In this study, we demonstrated the existence of mesoscopic two-dimensional (2D) coherent charge transport in poly(3,4-ethylenedioxythiophene):poly-(styrenesulfonate) (PEDOT:PSS) film by performing structural investigations and high-field magnetoconductance (MC) measurements in magnetic fields of up to 15 T. We succeeded in observing marked positive MCs reflecting 2D electronic states in a conventional drop-cast film. This low-dimensional feature is surprising, since PEDOT:PSS—a mixture of two different polymers—seems to be significantly different from crystalline 2D materials in the viewpoint of the structural inhomogeneity, especially in popular drop-cast thick films. The results of the structural experiments suggest that such 2D transport originates from the nanometer-scale self-assembled laminated structure, which is composed of PEDOT nanocrystals wrapped by insulating sheets consisting of amorphous PSSs. These results indicate that charge transport in the PEDOT:PSS film can be divided into two regimes: mesoscopic 2D coherent tunneling and macroscopic three-dimensional hopping among 2D states. Our findings elucidate the hieratical nature of charge transport in the PEDOT:PSS film, which could provide new insight into a recent engineering concern, i.e., the reduction of the out-of-plane conductance.**

Conductive polymers are becoming indispensable in both scientific research and practical applications because of their advantageous features, such as mechanical flexibility, low toxicity, and the potential for use in low-cost device fabrication via printing processes. In particular, poly(3,4-ethylenedioxythiophene):poly(styrenesulfonate) (PEDOT:PSS, which is produced by doping PEDOT with PSS) is one of the most commonly used polymers owing to its beneficial electric and thermoelectric properties, good stability under ambient conditions, and visible-light transmittance in thin-film form. [1–8] Electrical conductivities exceeding 1000 S/cm have routinely been achieved through the development of efficient methods, in particular, the addition of polar solvents to aqueous PEDOT:PSS solutions; [9–11] however, the charge transport mechanism is still not well understood in spite of its widespread utilization. The electrical properties of PEDOT:PSS films are strongly related to their complicated macromolecular structures; hence, clarifying the correlations between their electrical and structural properties is an active area of investigation in polymer science.

The anisotropic conductance, namely, the reduction of out-of-plane conductivity, in highly conductive PEDOT:PSS films is one of the crucial issues that must be unraveled for future applications. [12–17] Previous studies have indicated that the out-of-plane conductivity tends to be significantly less than the in-plane conductivity. Nardes et al. [12] reported an anisotropy factor of 500 in a spin-coated thin film, while Na et al. [13] found that this factor can dramatically increase up to 30,000 depending on the film morphology and molecular conformations. Liu et al. [14] and Wei et al. [15] recently obtained anisotropy factors of 3 and 7, respectively, in their thermal conductivity measurements. Importantly, these anisotropic features were observed even in thick (micrometer-scale) films, which seems to be inconsistent with the amorphous nature of polymers. This situation strongly suggests that structural self-assembly and the associated characteristic charge transport occur in these films.

In this study, we confirmed the existence of mesoscopic two-dimensional (2D) coherent charge transport in a commonplace PEDOT:PSS thick film owing to its self-assembled laminated structure. Positive magnetoconductance (MC), which reflects 2D weak localization (WL), was observed in a conventional drop-cast PEDOT:PSS thick film. We investigated the nature of this 2D charge transport by examining the structural aspects of the film by employing grazing incidence wide-angle X-ray scattering (GIWAXS) with synchrotron radiation. The film was determined to have a self-assembled laminated structure: conducting PEDOT nanocrystals wrapped by insulating sheets consisting of amorphous PSSs. These results indicate that 2D quantum phenomena can be observed even in commonplace polymer films and, furthermore, enabled the development of a new aspect of charge transport in PEDOT:PSS film: three-dimensional (3D) hopping among disorderly stacked mesoscopic 2D structures. This aspect well represents the anisotropic features observed in PEDOT:PSS films.

The experiments were performed using free-standing, ~5-μm-thick PEDOT:PSS films, as displayed in **Figure 1**a, which were fabricated by drop-casting (see Supporting Information for details). Figure 1b illustrates the temperature dependence of the electrical conductivity in a 0 T magnetic field ($\sigma_0$). Owing to the ethylene glycol (EG) doping, the conductivity is three times that of the pristine film. The conductivity at room temperature ($T = 285$ K) is 714 S/cm and decreases with decreasing temperature. The rate of decrease is quite small; hence, the conductivity of 190 S/cm at the lowest experimental temperature of 1.3 K is only approximately one-quarter of that at room temperature. The conductivity seems to remain finite as $T \to 0$, as shown in the inset of Figure 1b. Importantly, these behaviors are quite different from those exhibited by typical insulators at low temperatures. Thus, they strongly suggest that the electronic ground state of the EG-doped highly conductive film was in the "metallic" regime in terms of the disorder-induced metal–insulator transition. [18,19]

To investigate the charge transport in the highly conductive film in detail, we performed MC measurements for both transverse and longitudinal geometries; hereafter, we define MCs as $\Delta\sigma_{T,L}(H,T) \equiv \sigma_{T,L}(H,T) - \sigma_0(T)$, where $\sigma_{T,L}$ is the conductivity in a magnetic field $H$ and subscripts $T$ and $L$ denote transverse and longitudinal geometry, respectively. Remarkable positive MCs are observable in the transverse geometry results in the upper panel in Figure 1c. We should note that the positive MC was not observed in the pristine film (Figure S3, Supporting Information). Such positive MCs (i.e., negative magnetoresistances), originating from breaking of the constructive quantum interference of back-scattered waves, [20] provide strong evidence that our film was in the WL regime. [21,22] The MC intensities increase with decreasing temperature above 20 K and then begin to be suppressed in high magnetic fields. This suppression, namely, the negative contribution, rapidly increases with decreasing temperature; hence, the MCs are negative in almost all of the magnetic fields below 5 K (lower panel in Figure 1c). Notably, positive contributions are not evident in the longitudinal geometry results (upper panel in Figure 1d). In general, applied magnetic fields modulate both the orbital motions and spin polarizations of electrons. The former, that is, the WL effect, is only observable in the transverse geometry, whereas the latter is observable in both geometries. [20,23] Thus, the WL contribution to the MC can be extracted from the difference between $\Delta\sigma_T$ and $\Delta\sigma_L$.

**Figure 2**a shows the extracted WL contribution defined by $\Delta\sigma_{WL}(H,T) \equiv \Delta\sigma_T(H,T) - \Delta\sigma_L(H,T)$ at temperatures from 0.5 K to 100 K. Positive MCs are clearly observable. As the temperature decreases, the shape of the MC versus the magnetic field $H$ changes from bowl-like (with a slope of zero as $H \to 0$) to cusp-like. This temperature dependence is the behavior typically observed in 2D WL systems. [24] MCs owing to 2D WL can be described as follows:

$$\Delta\sigma_{WL} = A \frac{e^2}{2\pi^2 \hbar d}\left[\Psi\left\{\frac{1}{2} + \left(\frac{L_H}{L_\phi}\right)^2\right\} - \ln\left\{\left(\frac{L_H}{L_\phi}\right)^2\right\}\right] + B|H|$$

where $\psi(x)$ is the digamma function, $L_H = \sqrt{\hbar c/eH}$ is the magnetic length, $d$ is the film thickness, $L_\phi$ is the inelastic scattering length, and $A$ and $B$ are phenomenological parameters. The first term is the simplified Hikami–Larkin–Nagaoka equation, [24] and the second term, a linear background, originates from the inhomogeneous nature of the system, e.g., scattering from the grain boundary during hopping conduction between localized states. [25] The physical meanings of the fitting parameters are follows: $L_\phi$ is the average distance a charge travels between inelastic collisions, that is, a charge can propagate this distance while maintaining phase coherence; $A$ reflects the number of conduction channels through the film; and $B$ indicates the significance of the hopping effects. Notably, double-dip shapes are evident in the high-temperature region above 20 K (e.g., the pink curve for 100 K shown in Figure 2a). These might have originated from the anti-WL effects due to the spin-orbit interaction; [20] however, these effects were too small to separate from the significant WL effects. Thus, the fittings above 20 K were performed based on the results obtained using high magnetic fields (>8 T) to exclude anti-WL effects.

As shown by the gray broken curves in Figure 2a, the observed MCs are well represented by the above-mentioned theoretical model. The characteristic change of the curve shapes from bowl-like to cusp-like with decreasing temperature is associated with the increase of $L_\phi$. These results strongly indicate that 2D coherent transport with WL was realized in our film. As shown in Figure 2b, $L_\phi$ is 3.5 nm at 100 K and increases with decreasing temperature. Considering that a PEDOT nanocrystal typically has dimensions of a few nanometers, a 2D state is as the result of the overlapping of wave functions. In addition, $L_\phi$ is closely following the power law $L_\phi \propto T^{-\frac{p}{2}}$ where $p$ is the parameter reflecting the dominant decoherence (inelastic collision) mechanism. The values of $p$ can be divided into two categories: $p \sim 1$ (from 2 K to 20 K) and $p \sim 2$ (below 2 K). Specifically, $p \sim 1$ and $p \sim 2$ are associated with the situations in which phase-breaking occurs due to electron–electron (e-e) interactions [26] and electron–phonon (e-p) interactions, [27,28] respectively. That is, the observed behavior might indicate a change in the main scattering process. The phase coherence of a charge would mainly be disturbed by e-e interactions above 2 K and e-p interactions below 2 K. $A$ and $B$ will be discussed after the structural features. It should be noted that this situation, e-p interactions become apparent at low temperatures, is a peculiar behavior compared with the general interpretation—e-p scatterings would be suppressed due to the suppression of the thermal excitation of phonons. We consider that this characteristic feature might originate from the boundary scatterings due to the mesoscopic structure that will be discussed later. These scatterings play an important role only in the case when a charge could travel enough long distance with keeping the phase coherence; therefore, owing to the competition between the suppression of phonon excitations and growing importance of boundary scatterings, e-p interactions seem to become important in response to the increasing of $L_\phi$ at low temperatures. Of course, this interpretation should not exclude the possibility of that further studies could investigate that e-e interactions become dominant again at extremely low temperatures. We mention that the phase breaking time $\tau_\phi$, which related to the inelastic scattering length $L_\phi$ as $L_\phi \equiv \sqrt{\tau_\phi D}$ where $D$ is the diffusion coefficient, might shed more light on the WL physics. For example, the observed unique crossover of the dephasing mechanism can be discussed more deeply with $\tau_\phi$. However, we could not evaluate the meaningful value of diffusion coefficient by Hall measurements because of the hopping nature of the system. It is expected that future progress on the mobility measurements in PEDOT:PSS might provide the deep understanding of the WL physics and electronic coherence at low temperatures.

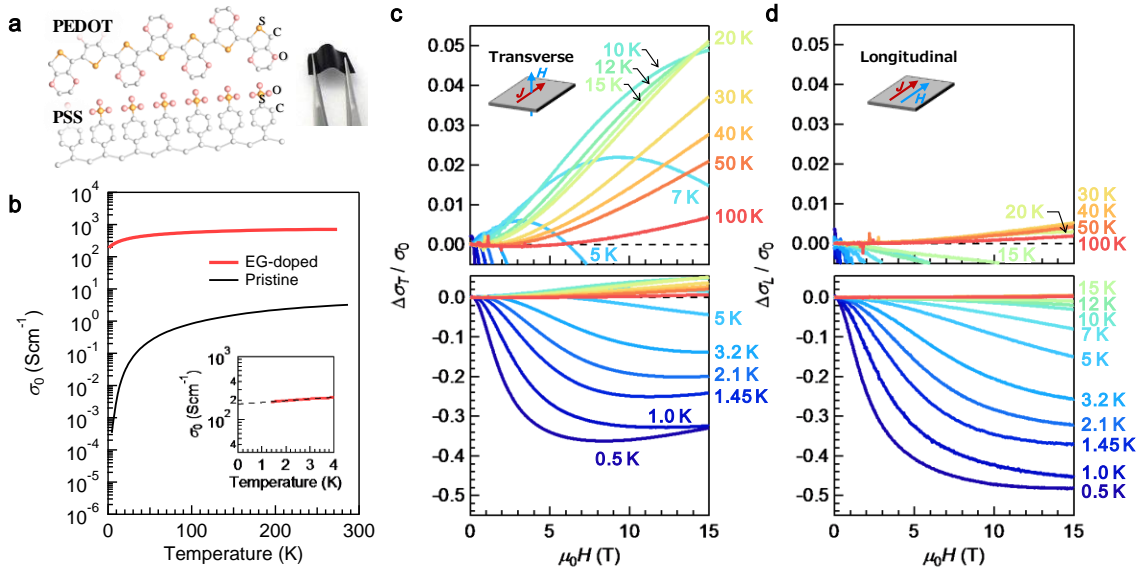

Figure 1. a) Chemical structures of PEDOT and PSS molecules and a picture of the free-standing film employed in this study. b) Temperature dependences of the electrical conductivities of the EG-doped (red curve) and pristine (black curve) PEDOT:PSS films. The inset is an enlarged version of the low-temperature region. The broken line is a guide for the eyes to facilitate estimation of the finite value approached as T→0 K. c) Transverse and d) longitudinal MCs divided by $\sigma\_0$ at various temperatures (0.5‑100 K). The upper and lower panels depict the positive and negative contributions, respectively. The insets in the upper panels are schematic illustration of the experimental geometries. H and J denote the directions of the magnetic field and current flow, respectively.

Considering that the inelastic scattering length is significantly less than the film thickness of ~5 μm, the observed 2D coherent transport should be associated with electrons confined within some nanometer-scale structure in the film. Thus, we focus on the structural aspects as the origin of above-mentioned 2D electronic features. **Figure 3**a shows a GIWAXS image of the PEDOT:PSS thick film. $q_y$ and $q_z$ are vectors representing the in-plane and out-of-plane scattering components, respectively. In addition to the isotropic amorphous halo of randomly distributed PSS around $q_{y,z}$ = 12.5 nm$^{-1}$, [29] three evident anisotropic scattering features are observable: intense scattering at $q_z <$ 5 nm$^{-1}$, moderate scattering at $q_z$ = 18.3 nm$^{-1}$, and weak scattering at $q_y$ = 24.3 nm$^{-1}$. In particular, the anisotropy of the small-$q$ scattering is quite significant. Figure 3b presents the scattering components $q_z$ (red curve) and $q_y$ (blue curve) extracted from Figure 3a. Intense scattering below 5 nm$^{-1}$ is observable only in $q_z$, which exhibits a peak at 2.1 nm$^{-1}$ and subsequent scattering with some small shoulders. The subsequent scattering is very weak compared with the main peak and was difficult to separate into individual peaks. Hence, we only focused on the peak at 2.1 nm$^{-1}$. The period of this peak was found to be 3.0 nm, which is sufficiently longer than all of the bond lengths and stacking periods of the PEDOT and PSS molecules. Therefore, the quite strong peak at 2.1 nm$^{-1}$ only in $q_z$ indicates large-scale vertical stacking with a period of 3 nm. Considering that PEDOT nanocrystals cling onto PSS molecules, [29] the expected, simplest, and most consistent picture is a laminated structure, which indicates alternate stacking of the PEDOT nanocrystals and PSS molecules.

The other anisotropic features at 18.3 nm$^{-1}$ and 24.3 nm$^{-1}$ provide more information about the anisotropic structure. The 18.3 nm$^{-1}$ peak corresponding to the $\pi$–$\pi$ stacking (0.343 nm period) of the PEDOT molecules is quite significant in $q_z$. [29–32] Furthermore, the integrated intensity of the $q_z$ peak was found to be roughly three times stronger than that of the $q_y$ peak (Figure S4, Supporting Information). This finding indicates that most of the PEDOT nanocrystals were oriented face-on in the film. This stacking feature is consistent with that observed in a previous study of PEDOT:PSS printed films. [33] Moreover, the 24.3 nm$^{-1}$ peak (i.e., 0.259 nm period) observed only in $q_y$ corresponds well to the dimensions of a PSS monomer along the chain, indicating the alignment of the PSS chains parallel to $q_y$.

Figure 3c provides a schematic illustration of the film structure identified based on the above-mentioned findings. For clarity, the amorphous and randomly oriented molecules are not drawn. The PSS chains lie parallel and are stacked perpendicular to the substrate. This stacking forms a laminated structure with a characteristic period of 3.0 nm.

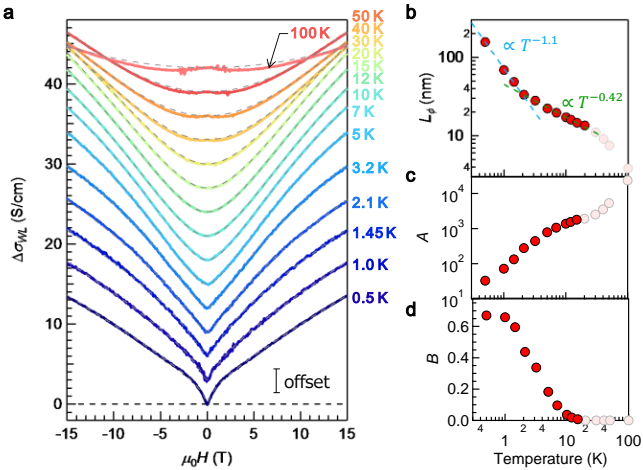

Figure 2. a) MCs reflecting 2D WL at various temperatures (0.5–100 K). The gray broken curves are fitting curves. Temperature dependences of b) $L_\phi$, c) $A$, and d) $B$. The light-colored points (above 20 K) indicate that the fits were performed based on the results obtained in high magnetic fields (>8 T). The broken lines in b) indicate the power-law dependences on the temperature.

The PEDOT nanocrystals cling to the PSS molecules and prefer uniaxial (face-on) orientation. This model is consistent with those identified in previous studies [33–35] and is in good agreement with the results of other experiments (e.g., an analysis of the lattice contributions to the anisotropic thermoelectric effects and a morphological study involving atomic force and scanning electron microscopy).[9,27]

Importantly, if a charge is confined within a thin layer having a thickness less than $L_\phi$, the charge transport becomes 2D from the viewpoint of not only the absolute conductivity but also the electronic coherence. Considering the $L_\phi$ results depicted in Figure 2b, this condition should be satisfied at least below 100 K; that is, the results strongly suggest that the observed 2D transport reflects the motion of electrons confined between intertwined PSSs. Of course, the area of this effective 2D structure would be significantly smaller than that of the bulk film, as shown in the schematic illustration (Figure 3c). PSS boundaries also exist in the in-plane direction; however, these are dilute compared to those in the out-of-plane direction. The mean distance between boundaries might be given by the above-mentioned change of the scattering mechanism from e-e (above 2 K) to e-p (below 2 K) interactions. The phase-breaking due to e-p interactions would play an important role only if $L_\phi$ were longer than the mean distance between PSS boundaries. Therefore, in our situation, ~30 nm could correspond to this distance, which is the in-plane length scale of the mesoscopic 2D structure.

The bulk transport through the film could be described as incoherent 3D hopping among such 2D structures. Importantly, the possible charge-hopping routes would be reduced with decreasing temperature because of the thermal energy reduction. Thus, the number of effective conduction channels through the film would decrease with decreasing temperature, which could be the physical meaning of the temperature dependence of $A$ (Figure. 2c). In addition, the dependence of $B$ on the temperature (Figure 2d) could indicate the relative importance of the grain boundary effects. The high values of $B$ at low temperatures might be associated with the fact that scattering from the grain boundary becomes significant at such temperatures.

In conclusion, we elucidated the existence of hierarchical charge transport in PEDOT:PSS film, that is, "island-hopping" among mesoscopic 2D structures. The results indicate that this self-assembled mesoscopic structure, consisting of PEDOT nanocrystals confined by amorphous PSS sheets, could have a thickness of ~3 nm and an effective horizontal scale of ~30 nm. A charge could travel a long distance with little scattering within this structure and then hop to another 2D structure. Hence, the in-plane transport would be enhanced compared to the out-of-plane transport. Importantly, in the low-temperature region in which $L_\phi$ is larger than the structure thickness ($T < 100$ K), we succeeded in observing the effects of 2D WL on the MC. The results indicate that 2D coherent transport can be realized even in commonplace polymer films fabricated by simple drop-casting using a popular material, rather than only in highly crystalline materials.[37,38] Our findings could provide not only new insight into the conductance anisotropy in PEDOT:PSS, but also the ability to explore a wealth of 2D quantum physics phenomena hidden in polymers.

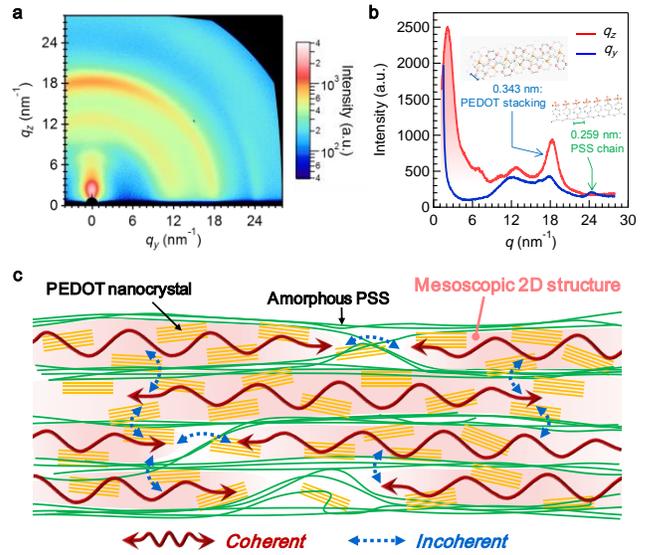

Figure 3. a) GIWAXS image of a PEDOT:PSS thick film. b) Scattering profiles of $q_z$ (out-of-plane) and $q_y$ (in-plane). The inset illustrations show the structures of a PEDOT nanocrystal (upper) and PSS molecule (lower). c) Simplified cross-sectional illustration of the laminated structure. The pink regions indicate mesoscopic 2D structures. The red and blue arrows denote coherent and incoherent charge transport paths, respectively.

**Acknowledgements**

The synchrotron radiation experiments were performed at BL40B2 of SPring-8 with the approval of JASRI (proposal nos. 2015A1773, 2015B1694, and 2016A1617). We thank H. Sekiguchi and N. Ohta for the technical assistance in these experiments. We also thank Y. Kato, N. Asano, and K. Hashimoto for the fruitful discussions. This work was supported by JSPS KAKENHI grant numbers JP15K17688 and 16K05430 and the Foundation for Promotion of Material Science and Technology of Japan.

# Supporting Information

**Mesoscopic 2D Charge Transport in Commonplace PEDOT:PSS Films**
Yuta Honma, Keisuke Itoh*, Hiroyasu Masunaga, Akihiko Fujiwara, Terukazu Nishizaki, Satoshi Iguchi, and Takahiko Sasaki

## I. EXPERIMENTAL METHODS

***Film Preparation.*** The PEDOT:PSS thick films were prepared via drop-casting. An aqueous solution (Clevios™ PH 1000) with an additional solvent (EG 3 wt.%) was dropped onto silicon substrates (10 mm × 10 mm × 0.525 mm). The film called "pristine" in the manuscript was fabricated in the same way but without an additional solvent. The films were dried on a hot plate at 100 °C for 10 min and 160 °C for 10 min and were ~5 μm thick.

***X-Ray Scattering Measurements at SPring-8.*** The GIWAXS measurements were performed at BL40B2 of SPring-8 in Hyogo, Japan. The incoming X-ray wavelength and incident angle were set to 1 Å and 0.15°, respectively. This incident angle was larger than the critical angle. The distance between the sample and detector was 293 mm and was calibrated using a reference sample of silver behenate.

***Transport Measurements.*** The electrical conductivity was measured using a four-probe method along the in-plane direction with the voltage electrodes separated by 0.5 mm. The MC measurements were performed using a superconducting magnet (Oxford Instruments) to create magnetic fields of up to 15 T. To obtain the low temperatures down to 0.5 K, we used the $^3$He insert with a pumping system in addition to the standard variable temperature insert of the superconducting magnet.

## II. REDUCED ACTIVATION ENERGY

Figure S1 shows the reduced activation energy $W = d(\ln \sigma_0)/d(\ln T)$ as a function of the temperature (a so-called Zabrodskii plot), which is known to be useful for differentiating among the "metallic," "critical," and "insulating" regimes in the context of the metal–insulator transition in conducting polymers.[S1,S2] It can be seen that our film was in the "metallic" regime because $W$ is not constant, but rather increases with increasing temperature in the low-temperature region.

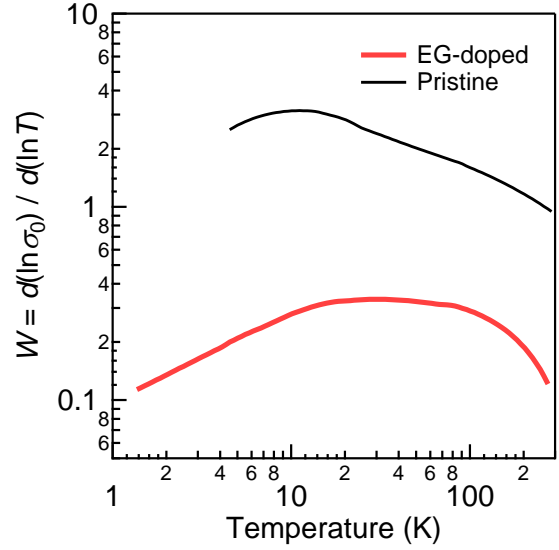

**Figure S1.** Temperature dependence of $W = d(\ln \sigma_0)/d(\ln T)$ of the EG-doped (red curve) and pristine (black curve) PEDOT:PSS films.

## III. BULK CHARGE TRANSPORT

As shown in Figure S2a, the temperature dependence of the electrical conductivity in the pristine film was well represented by the exponential function ($\sigma(T) \propto \exp\left\{-\left(\frac{T_0}{T}\right)^\gamma\right\}$, where $T_0$ is the characteristic temperature and $\gamma$ is the parameter reflecting the hopping process and dimensionality). Such exponential-type temperature dependence and the value of $\gamma \sim 0.5$ indicates that the charge transport in the pristine film can be understood by the Efros-Shklovski variable-range hopping (VRH).[S3] In contrast, the EG-doped film does not show exponential-type dependence. As shown in Figure S2b, the temperature dependence of the electrical conductivity followed the power-law type dependence ($\sigma(T) \propto T^{1/2}$) at low temperatures.[S2,S4] Such changes in the temperature dependence, the exponential-type to the power-law type, indicates that tunneling process is more significant than the hopping process in the EG-doped film.

We should note that the power-law of $T^{1/2}$ in Figure S2b corresponds to the 3D WL. An ideal 2D system should follow the logarithmic dependence of $\sigma(T) \propto \log T$.[S4] This discrepancy of the dimensionality between the MC and bulk charge transport might indicate that the 2D transport is limited within the mesoscopic structure.

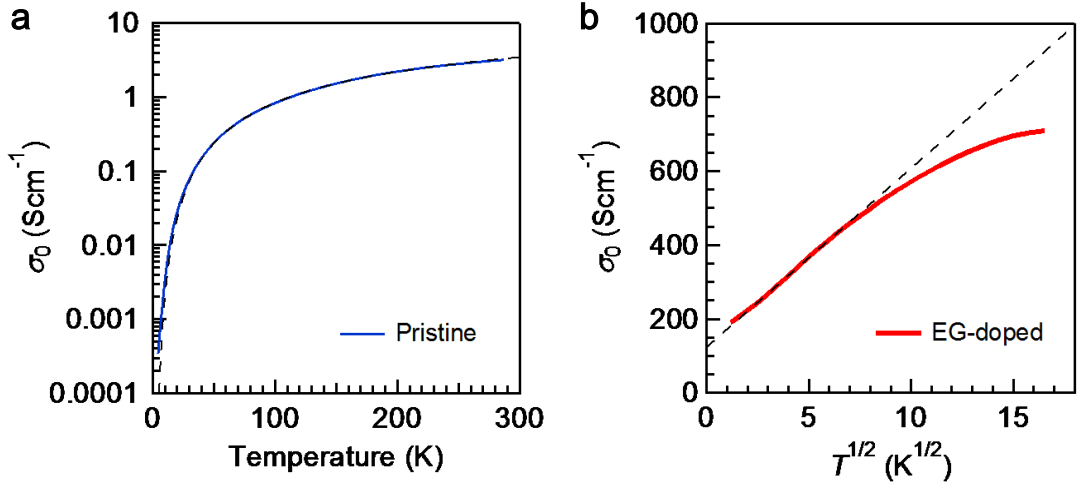

**Figure S2.** The temperature dependence of the electrical conductivity in the a) pristine film and b) EG-doped film to obtain the transport model. The black broken curves in a) and b) indicate the exponential ($\sigma_0(T) = A \exp\{-\left(\frac{T_0}{T}\right)^\gamma\}$) and power-law ($\sigma_0(T) = B + CT^{1/2}$) dependence, respectively. Numerical fittings were performed at the temperature ranges of 10-250 K in a) and 4-50 K in b). Obtained values were $A = 44.52$, $T_0 = 3097$, $\gamma = 0.4011$, $B = 124.0$, and $C = 48.41$.

## IV. MAGNETOCONDUCTANCE IN PRISTINE FILMS

In contrast to our study, no positive MC was observed in the spin-coated PEDOT:PSS films doped with dimethyl sulfoxide (DMSO).[S5] One possible origin which causes this difference would be the difference of the leading transport mechanism in the film. The temperature dependence of the electrical conductivity in that film was well represented by the VRH transport; it indicates that their film is still in the "insulating" regime in contrast to our "metallic" film. In such case, the positive MCs reflecting the WL nature could not exist or be concealed by the negative MCs originating from the hopping nature. It is noteworthy that both this film (DMSO-doped film) and our EG-doped film have the similar high conductivity at room temperature; however, the temperature dependence is quite different. Further studies are needed in order to investigate the origin of this difference.

To elucidate the MC behavior in the "insulating" PEDOT:PSS film, we performed MC measurements in the pristine film. Figure S3 shows transverse MCs observed in the pristine (black curve) and EG-doped films (red curve) at 20 K. The positive MC was not observed in the pristine film which exhibits the insulating behavior (see Figure 1b). This result indicates that the positive MC reflecting the WL in the PEDOT:PSS system could be observed only in the "metallic" film.

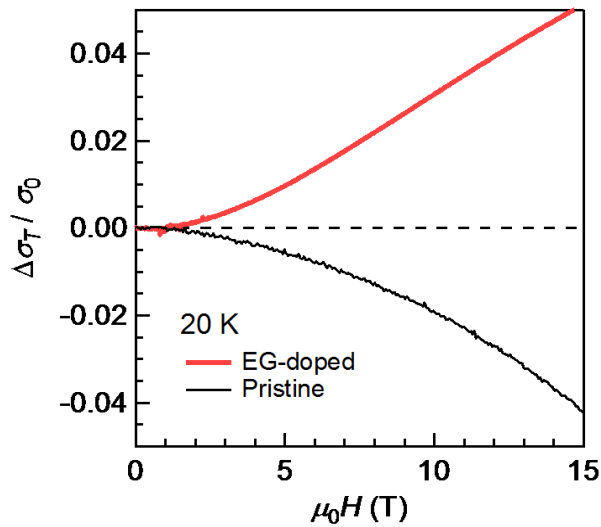

**Figure S3.** Transverse MCs observed in the EG-doped (red curve) and pristine (black curve) PEDOT:PSS films at 20 K.

We should mention that since the difference of the characteristic length scale between MC and bulk transports, the VRH temperature dependence of the electrical conductivity does not completely exclude the possibility of the WL; it is not enough to exclude the existence of the micro- or mesoscopic WL nature. The electrical transport through the bulk polymer materials strongly depend on the inter-domain transport; thus, even though the system contains metallic domains, an insulating behavior sometimes appears due to the inter-domain hopping transport. The previous study on the doped PBTTT (poly(2,5-bis(3-hexadecylthiophen-2-yl)thieno[3,2-b]thiophene) films[S5] might be an example of this situation; the positive MC originating from metallic domains coexists with the insulating bulk electrical conductivity (conductivity decreases with temperature decreases). It might indicate that the inter-domain transport of such PBTTT films was still in the hopping regime.

## V. PEAK INTEGRATION

Figure S4 is the enlarged figure of Figure 3b to estimate the integrated intensity of the 18.3 nm$^{-1}$ peak. After subtracting a linear background between 14.5 nm$^{-1}$ and 22.5 nm$^{-1}$ (black dotted line in Figure S4), the numerically integrated intensity in $q_y$ and $q_z$ are 465.2 (light blue hatch) and 1402 (light red hatch), respectively.

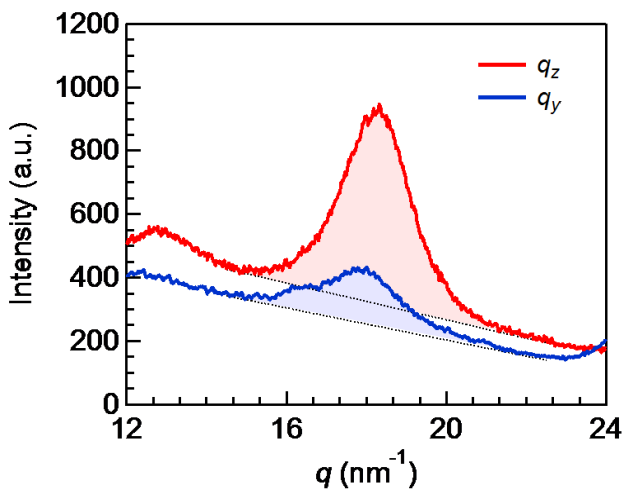

**Figure S4.** The enlarged figure of Figure 3(b) around the 18.3 nm$^{-1}$ peak (PEDOT π-π stacking).

## VI. GIWAXS ANALYSIS

The GIWAXS analysis whose results are displayed in Figure 3(a), 3(b) and S4 was based on the ($q_y$, $q_z$) coordinates. The region missing from around the $q_z$ axis due to the existence of $q_x$ was not considered.[S6] Figure S5 shows an accurate image using ($q_r$, $q_z$) coordinates. The pole-figure of the PEDOT π–π stacking peak at $q = 18.3$ nm$^{-1}$ (black curve) and peak interpolation by a Lorentzian function (red broken curve) are depicted in Figure S6. Quantitative analysis could not be performed because of the absence of specular diffraction and the existence of significant background scattering; however, marked anisotropy is evident distribution around the rocking angle $\chi = 0°$.

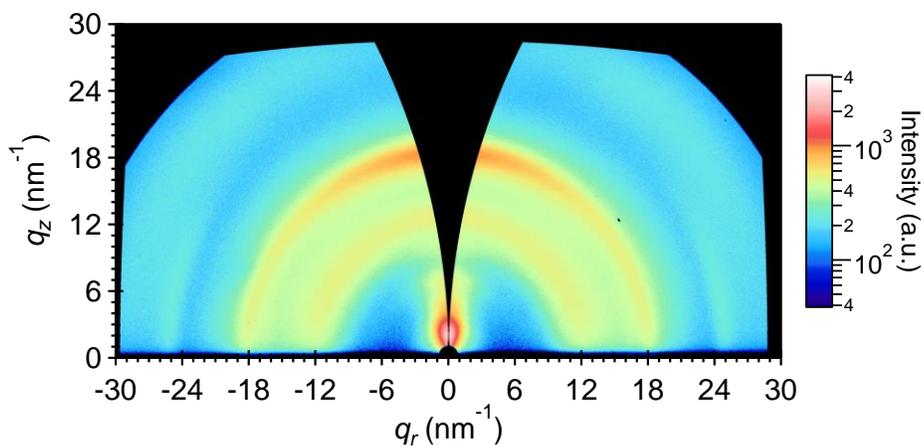

**Figure S5.** GIWAXS image plotted using ($q_r$, $q_z$) coordinates.

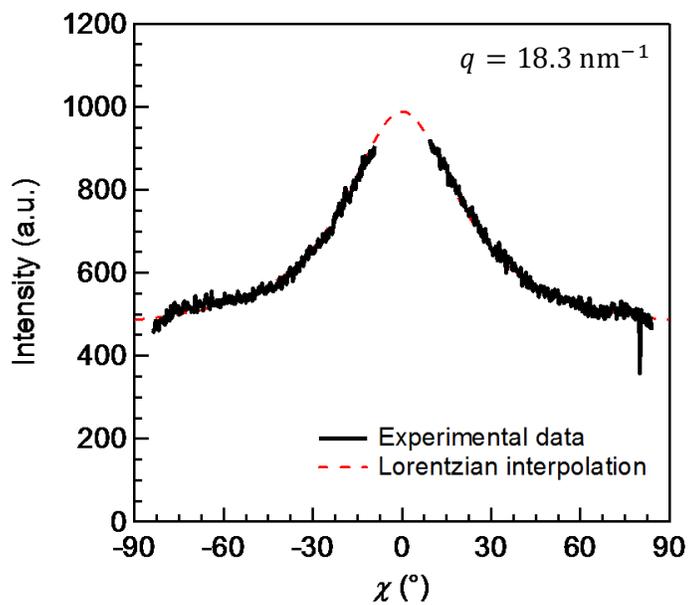

**Figure S6.** Pole-figure of the PEDOT π–π stacking peak intensity. The black and red broken curves are the experimental data and Lorentzian interpolation, respectively. The full width at half-maximum is 47.7°.